# Zero Echo Time Functional MRI in Humans

Ali Caglar Özen[1], Shuai Liu[1], Serhat Ilbey[1], Michael Bock[1], Uzay Emir[2,3], Yen-Yu Ian Shih[4],

[1]Division of Medical Physics, Department of Radiology University Medical Center Freiburg, Faculty of Medicine, University of Freiburg, Freiburg, Germany
[2] Department of Radiology, University of North Carolina, Chapel Hill, NC, USA.
3 The Lampe Joint Department of Biomedical Engineering, University of North Carolina, Chapel Hill, NC, USA.

[4] Department of Neurology, University of North Carolina, Chapel Hill, NC, USA.

Short running title: ZTE fMRI in humans

Word count: 5120
Number of Figures: 8
Number of Tables: 0
Number of References: 104

**Keyword(s):** Zero Echo Time, fMRI, BOLD, resting state, UTE, ZTE, PETRA

## Abstract

Conventional EPI-based fMRI relies on the BOLD contrast to map activity changes, but is limited by high acoustic noise, susceptibility artifacts, and distortion near air-tissue interfaces. Ultra-short and zero-echo-time (ZTE) sequences have gained considerable attention in animal fMRI studies due to their robustness against susceptibility effects, low acoustic noise, and ability to capture signal sources distinct from those of conventional BOLD imaging. In this study, we demonstrate the feasibility of using ZTE for fMRI applications in humans at 3T, while maintaining many of the advantages observed in preclinical studies. Phantom measurements confirmed that the ZTE protocol is sensitive to oxygen concentration changes at 3T, supporting its potential for detecting related functional signals. In five healthy volunteers, task-based visual stimulation and finger tapping experiments elicited robust activation in the primary visual cortex and motor cortex, respectively, using the ZTE sequence. Resting-state networks were successfully identified using ICA. Compared to BOLD-EPI, ZTE showed significantly reduced acoustic noise and was largely insensitive to susceptibility-induced signal loss. While this work represents our initial experience using ZTE for human fMRI at 3T, the presented results will motivate future studies to refine and extend this approach to explore complementary functional contrast mechanisms.

200 words

# Introduction

Functional magnetic resonance imaging (fMRI) with the blood oxygen level-dependent (BOLD) contrast has become an indispensable tool for mapping brain activity[1–3]. Conventionally, BOLD fMRI is realized by echo-planar imaging (EPI) sequences[4]. Variants of EPI have proven to offer distinct advantages and challenges[5–12], highlighting the importance of understanding the signal's origins and optimizing acquisition parameters. These challenges stem from the intertwined nature of neuronal, metabolic, and vascular processes, which are further influenced by various instrumental and physiological factors[13]. Common EPI problems include high sensitivity to susceptibility artifacts, limited temporal resolution, challenges related to eddy current-induced distortions, and the generation of high acoustic noise. Complementary MRI methods for detecting changes in brain activity could provide insights into the complex mechanisms underlying task-related activation and resting-state functional connectivity. Free induction decay, i.e., FID-based pulse sequences similar to RUFIS (also referred to as ZTE), MB-SWIFT, UTE, PETRA[14–20] might also offer complementary solutions for addressing common EPI challenges in studies where susceptibility, distortion, or acoustic noise represent significant limitations.

The premise of BOLD fMRI relies on the coupling between neuronal activity and subsequent hemodynamic changes, including cerebral blood flow (CBF), cerebral blood volume (CBV), and cerebral metabolic rate of oxygen (CMRO2). A temporal resolution of 1 or 2 seconds has proven useful in capturing these mesoscale hemodynamic changes. While BOLD remains the primary approach due to its simplicity and broad applicability, alternative fMRI methods have been proposed to enable more direct observations of neuronal activation through MRI[21–24]. However, each of these methods faces its own challenges. Diffusion-weighted fMRI can be confounded by BOLD effect[25,26], peripheral nerve stimulation (PNS), and specific absorption rate (SAR) considerations in spin-echo EPI, especially at ultra-high fields, making measurements with sub-second temporal resolution challenging. Non-EPI methods might be useful here to suppress BOLD effects and PNS. Recently, a rapid onset of diffusion MRI signal was reported in a preclinical setting[27], which has contributed to the understanding and interpretation of direct detection of neuronal activity. However, the mechanisms contributing to this signal are still an active area of research[28–30]. Functional MR spectroscopy, a technique that detects activation-related variations in neurochemical signals, is also subject to experimental considerations such as SAR and signal-to-noise ratio (SNR), which limit its temporal resolution[31,32]. Elastography-based functional MRI offers access to high-frequency neuronal activity; however, its model-based reconstruction relies on spatial derivatives of the signal, which makes it challenging to achieve a high degree of spatial specificity[33].

Although no technique has displaced GRE-EPI as the gold standard, advances have been made to address its limitations. Techniques such as arterial spin labeling (ASL) [34–36], the spin-echo (SE) variant of EPI[37,38], and vascular space occupancy (VASO)-fMRI[34,39,40] provide more spatially localized functional sources than GRE-BOLD-fMRI. However, because these techniques still rely on EPI readouts out of necessity for rapid acquisition, they suffer from many of the disadvantages of conventional EPI-fMRI, such as artifacts from ghosting, motion, and sensitivity to magnetic field inhomogeneities. To overcome acoustic noise problems in fMRI, a looping star fMRI sequence has been implemented[41–44]. However, due to the refocusing of gradient echoes to achieve adequate T2* sensitivity to BOLD signal changes, the sequence is sensitive to field inhomogeneities and lacks signal preservation in areas such as the orbitofrontal cortex and temporal lobe. Recently, a multi-echo implementation of the looping star fMRI was applied to sleep research[45]. Another recent work addressed the aliasing problem to improve spatial resolution using temporal filtering and model-based reconstruction[46,47]. Looping star was also successfully demonstrated at 7T[48]. Although the looping star is able to acquire an FID, as in conventional ZTE, it is a dual echo acquisition; and so far only the BOLD-weighted second echo signals have been analyzed, where, the focus was not to amplify T1-related signal changes as in the preclinical ZTE and SORDINO fMRI studies[49–51]. Multimodal approaches combining ASL, VASO, and BOLD show promise for quantifying neurovascular coupling parameters, such as oxygen extraction fraction (OEF) and cerebral metabolic rate of oxygen ($CMRO_2$), but require further optimization for clinical translation. Advances in sequence design (e.g., 3D-GRASE VASO, short-PLD ASL) and high-field MRI hardware may mitigate current limitations[35,36,40]. While techniques beyond our brief summary here have overcome drawbacks of GRE-EPI, no existing human fMRI technology is capable of addressing all of these problems. We believe ZTE-fMRI has the potential to overcome the most notable limitations of GRE-EPI.

Challenges in conventional EPI-based sequences mainly include acoustic noise, artifacts, and distortion. Specifically, acoustic noise from EPI can induce hearing loss and stress[52] , and confound fMRI results in both task-based and resting-state conditions[53]. In standard practice, the 2D EPI sequence collects multiple slices to obtain isotropic resolution. To accommodate these slices within brief repetition times (TRs), conventional protocols often compress the acquisition, which significantly elevates sound pressure levels (SPL). Furthermore, the Lorentz forces responsible for acoustic noise scale logarithmically with both the strength of the main magnetic field and the gradient field strength, resulting in more intense scanner noise at ultra-high field strengths, especially at high-resolution protocols[54]. Pioneering studies indicate that using silent fMRI sequences can improve data quality in studies involving working memory tasks[55], auditory stimuli experiments[56], and functional connectivity measurements[57]. To date, "silent" fMRI sequences have been reported to reduce SPL by 7.2–30 dB compared to standard EPI sequences[41–43,56], while sequences based on continuous gradient ramp can reduce SPL by as much as 40 dB[58–60]. Notably, techniques such as MB-SWIFT appear promising in suppressing acoustic noise while reducing motion confound[61], suggesting the utility of sampling data at ultra-short echo time (UTE) or zero echo time (ZTE). Emerging rodent studies have shown that ZTE-fMRI can detect functional activations through an endogenous contrast[49–51,60,62–64] with greater sensitivity than the BOLD fMRI with EPI[49]. Low flip angle ($\alpha<10°$) and short TR in ZTE result in a proton-density and T1-weighted contrast, which has been shown to be useful in capturing tissue oxygen-related T1-shortening as a surrogate marker of neural activity in rodents[49]. Despite these initial rodent findings, the feasibility of using ZTE for human fMRI has never been demonstrated due to challenges associated with long transmit/receive (Tx/Rx) switching dead times and the availability of a sequence that can acquire data at high temporal resolutions. The only relevant report the authors could find was the application of the SWIFT sequence to fMRI in humans and mice, demonstrating the sensitivity of the sequence to functional activation, which dates back to 2012 and has not yet been reproduced [65]. This study was done using small local Tx/Rx surface coil and therefore signal must be highly influenced by in-flow type contrast. Sensitivity of the magnetization-prepared ZTE to arterial blood volume changes has been studied using magnetization transfer pulses[66] in humans. Here, a data with magnetization transfer pulse set to 0° flip angle was also shown for a single-subject visual stimulus experiment, but only apparent activation could be observed in the sagittal sinus, most likely due to blood volume changes within the sinus as reported. ZTE has also been used for fMRI using a T2 preparation pulse to achieve BOLD contrast at a long echo time in looping star[67], i.e., using the second echo of a dual echo ZTE sequence. It is yet unknown whether tissue oxygenation-related contrast is measurable with the ZTE sequence at clinical field strengths. A recent review of UTE/ZTE-based fMRI studies also highlights that the translation of these techniques to clinical systems is currently a partially unfulfilled goal[68].

This study presents exciting empirical data with ZTE-fMRI in humans on a 3T clinical whole-body MRI system. We programmed the ZTE sequence suitable for fMRI, recruited five healthy volunteers for a standard visual task, evaluated the impact of the Tx/Rx switch dead-time, compared the effects of missing data points at the center of k-space, analyzed the evoked response, and tested the in-flow contributions on the functional data. A task-free measurement was also performed to compare the resting-state networks to conventional BOLD-EPI. As fMRI continues to advance at ultra-high magnetic fields, we anticipate that the ZTE fMRI sequences, which are free from EPI-related confounds and capable of capturing distinct physiological metrics, could further enhance the toolbox for functional brain mapping in humans.

# Methods

In ZTE, the readout gradient is already ramped up before the RF pulse[14,69,70], resulting in the first samples of each spoke being missed (Fig 1). For a higher maximum readout gradient, Gmax, the radius of the spherical gap ($k_{gap} = TE \cdot G_{max}$) increases. The total number of single-point imaging (SPI) samples required, $N_{SPI} =$ $4\pi(TE/t_d)^3/3$, $TE$ is the time between the center of the RF pulse and the acquisition of the first data point along a radial spoke$t_d$ its dwell time ($N_{SPI} \propto t_d^3$ ). ZTE was based on PETRA sequence[18,71], to fill in the sampling gap at the center of the k-space using SPI[72,73] (Fig. 1b). k-space view ordering follows a spiral phyllotaxis trajectory as shown in Supporting Information Video S1 for two subsequent volumes. To reduce the time required for SPI acquisition, compressed sensing PETRA implementation was used [74,75] as shown in Figure 1d. For convenience, all ZTE-based sequences are referred to as "ZTE" unless a specific comparison is made between different implementations. All sequences used in phantom and in vivo experiments, including tSNR measurements, were performed using the same in vivo EPI and ZTE fMRI protocols unless otherwise specified.

Dynamic 3D ZTE measurements were performed for five healthy volunteers (24-37-year-old, M) using the csPETRA sequence variant with the following parameters: TR/TE = 2/0.06 ms, $T_{enc}$ = 1.6 ms, $T_{RF}$ = 8 μs (Rectangular RF pulse, 10mT/m gradient strength during RF pulse, 3-dB-excitation-FOV = 192 mm), α=3°, FOV = (192 mm)$^3$, Matrix size = 64$^3$, ΔV = (3 mm)$^3$, $N_{spokes}$ = 1500, $T_{acq}$ = 3s/volume, $N_{SPI}$ = 27-single-point (SPI) acquisitions, 106 volume series, $T_{acq}$ = 5:21 min:s. TR and α were selected to maximize $T_1$-contrast based on Bloch-equation simulations (TR = 0.5:0.5:5 ms, α = 1:0.5:5°, $T_1$ = 1400 ms). Minimum RF pulse duration to achieve α = 3° was 8-μs-long rectangular pulse due to system's peak RF power limitations. For comparison, an EPI-BOLD measurement was performed (TR/TE = 3000/30ms, α=90°, $N_{slices}$=54, FOV=192x192x162mm$^3$, ΔV = (3mm)$^3$, 106 volumes series, $T_{acq}$ = 5:18 min:s). A $T_1$-MPRAGE (TR/TE/TI = 2000/2.41/900 ms, FOV=(192mm)$^3$, ΔV=1x1x1mm$^3$, α=8°) was acquired as anatomical reference. For all fMRI measurements, visual stimulation with a flickering checkerboard pattern and visually-triggered finger tapping task was applied with a 42 s=21 s (OFF) + 21 s (ON) block design paradigm (Supporting Information Fig. S1). The same protocols were used for resting-state measurements, without any task present. Last 10 images from resting state measurements were also used to calculate tSNR for two subjects (Subjects 1 & 2). All human imaging was carried out in accordance with relevant guidelines and regulations. Healthy volunteer scanning was approved by the Institutional Review Board of the University Medical Center Freiburg (No. 160/2000), and informed written consent was obtained prior to imaging.

The acoustic noise adjacent to the magnet and the additive acoustic noise of csPETRA were measured for $N_{spokes}$ = 100,000 and $N_{spokes}$ = 1,500 for various $G_{max}$ and Acc using a digital sound level meter (DSL 331, Tecpel Co., Ltd., Taiwan) and compared to EPI.

To investigate the influence of undersampling on activation sensitivity, ZTE-fMRI data were retrospectively reconstructed with only 50% of all spokes, resulting in a 2-fold increase in temporal resolution. The effect of the SPI samplings at the central portion of the k-space was also tested by performing the same analysis on the images reconstructed by omitting the 27 SPI samples, corresponding to 54 ms of additional acquisition per volume for Subjects 1 and 2. Unless otherwise stated, for the reconstruction of the ZTE images, radial projection data were first re-gridded onto a Cartesian grid using the Kaiser-Bessel convolution kernel and then combined with the SPI data after applying Hamming and density compensation filters. Finally, the image was divided by the apodization function, i.e., the Fourier transform of the gridding kernel. Data from each receive channel was reconstructed independently, and then the images were combined using the sum-of-squares method. The reconstruction of a 4D data set took approximately 12 min using a workstation with a 3.6-GHz six-core CPU and 128 GB RAM. For reconstruction of the T1-MPRAGE and GRE-EPI data, vendor-supplied automated image reconstruction software was used

In the volunteer data, a General Linear Model (GLM) analysis was performed, and T-scores were calculated (the FSL scripts used for post-processing and data analysis are provided in the Supplementary Material). The same post-processing steps were applied to both EPI and ZTE data. Motion correction did not result in any improvements for the ZTE data, as radial acquisition is robust against motion due to oversampling of the central k-space. Following the preclinical results[49,76], the same hemodynamic response function was assumed for the data acquired with both pulse sequences. T-scores were clustered with a threshold of $t \geq 2$ and mapped onto the $T_1$-MPRAGE image. Resting state networks were calculated using independent component analysis (ICA)[77] . For each ICA component, the voxel-wise intensity values in the spatial maps are parameter estimates (converted to z-scores) indicating the degree of synchronization between those voxels and the specific component/network, i.e., functional connectivity. Seed-based correlation maps were also obtained for the selected voxels with peak z-score and overlaid on the ICA results. For both task-based and resting state data, group ICA analyses were also performed after registering the original images to the MNI152 atlas.

To investigate the temporal SNR (tSNR) and to test td, TR, and Gmax parameters for subsequent in-vivo fMRI measurements, a homogenous cylindrical phantom (Volume = 1000 ml, $T_1$ = 451±48 ms, $T_2$ = 95±12 ms) was measured at a 3T clinical MRI system (Magnetom PrismaFit, Siemens, Erlangen) using a 20-channel head coil array (Head/Neck 20, Siemens). tSNR was calculated as $tSNR(x,y,z) = \frac{\mu}{\sigma}$ where $\mu = \frac{\sum_{i=1}^{N_I} I(x,y,z)}{N_I}$, $\sigma = \sqrt{\frac{\sum_{i=1}^{N}(I(x,y,z)-\mu)^2}{N_I}}$ for $N_I$ is the number of images and $I$ is the magnitude image. In addition to the in vivo dynamic measurement protocols given below, for TR = 2 ms and $G_{max}$ = 2.6 mT/m, dwell times of $t_d$ = 2 μs and $t_d$ = 8 μs were tested. For TR = 1 ms, $G_{max}$ of 5.2 mT/m and 10.4 mT/m were also tested. For $G_{max}$ = 10.4 mT/m, an acceleration factor of Acc = 16 was used to match the volume rate of acquisition to the other protocols[74]. Lower TR in ZTE enables acquisition at a higher volume per second. In this phantom experiment, TR values of 1 ms and 2 ms were compared. We also tested the effect of longer dwell times, which corresponds to a temporal averaging of sampled bits in the receiver, and compared $t_d$ = 8 μs to $t_d$ = 2 μs. Although it is expected that increasing the gradient strength will reduce the tSNR, this comparison was also measured.

To evaluate the dependence of the contrast-to-noise ratio (CNR) of the sequence on oxygen level, a set of phantoms was prepared with $O_2$- and $N_2$-saturated distilled water with 75 ppm contrast agent (Gd-DTPA, Magnevist, Bayer, Germany). Two 50-ml tubes were exposed to $O_2$ and $N_2$ gases via a 10-mm pore-size diffuser for 20 minutes. One tube was left as the control sample. Signal intensities were compared for the in vivo dynamic ZTE fMRI protocols given below. Note that the tubes were not pressurized; therefore, all the tubes were measured simultaneously. T1 values were also measured using a saturation recovery[78,79] TurboFLASH sequence (TR/TE = 5/2 ms, $\alpha$ = 8°, FOV = 192mm$^2$, $\Delta$V = 1x1x3 mm$^3$, BW = 390Hz/px) repeated for saturation times, $T_{SR}$ = {400:100:3000} ms. Here, saturation recovery-based T1 estimation was used. T2-SPACE images were acquired to suppress T1-weighting effects and to assess potential signal variations arising from coil sensitivity. Sequence protocols for both T1-mapping TurboFLASH and T2-SPACE are given in the Supplementary Material.

# Results

In Figure 2, activation maps for GRE-EPI and ZTE are presented, using the body coil as the transmit coil and the 20-channel head coil array as the receive coil, for the visual task. EPI shows activation diffused over a larger volume, whereas ZTE activation focuses on the right lingual gyrus. Activation maps for individual subjects, together with the time courses of the voxel with the highest Z-score, are provided. In Figure 3, activation map results from group analysis of both the visual and motor tasks for right-hand finger tapping are presented. Significant clusters were identified using a voxel-wise threshold of $p < 0.05$ ($|Z| > 1.64$) and a cluster-extent threshold of 80 voxels (FWER-corrected $p < 0.05$) in the group analysis of the visual task results (Figure 3). A single cluster centered in the posterior cingulate cortex (MNI: 16, -32, 52) survived correction (137 voxels, peak Z=8.07). Clusters with higher significance than the visual task were identified using a voxel-wise threshold of p < 0.05 ($|Z| > 3$) and a cluster-extent threshold of 300 voxels (FWER-corrected p < 0.05) in the group analysis of the motor task results. A single cluster centered in the left primary sensorimotor cortex—specifically the precentral gyrus (primary motor area M1) and postcentral gyrus survived correction (438 voxels, peak Z=8.79). Subject-specific Z-score maps are provided in Supporting Information Figure S2.

Figure 4 presents the primary visual network (PVN) and the sensory-motor network (SMN) derived from resting-state GRE-EPI and ZTE acquisitions. In general, the functional networks identified with GRE-EPI showed higher Z-scores compared to those from ZTE, consistent with findings reported in preclinical studies using similar pulse sequences[49,76]. For the PVN, both seed-based correlation analysis and ICA revealed consistent patterns, as illustrated in the overlay image. Group ICA of the ZTE data (five subjects) revealed several large and spatially distinct resting-state clusters. The largest cluster (5727 voxels, p < 0.01) corresponded to either the default mode network (DMN) or a major SMN. Additional clusters (3068 and 1253 voxels) also aligned with well-known resting-state networks, likely involving attention and visual regions, and thus classified as the PVN. Other prominent clusters (4674, 3976, and 1253 voxels, p < 0.01) were located in the parietal and temporal lobes, with one cluster (3976 voxels) possibly reflecting the DMN or an SMN. Overall, the size and spatial distribution of these clusters support the identification of canonical resting-state networks such as PVN and SMN.

Figure 5 shows activation maps following the visual stimulation overlaid on T1w-MPRAGE images in three orthogonal slices. The images were reconstructed with (PETRA) and without using the SPI samples at the k-space center (ZTE). Activation maps and resulting T-scores are in good agreement, suggesting that SPI acquisition is not essential for the low-resolution protocol and the specific experimental design used here. The Pearson correlation coefficient between the time courses of both dynamic volume series is r = 0.987, confirming a strong positive linear relationship. The acoustic noise measured next to the magnet without active imaging was 56 ± 1 dBA, whereas the acoustic noise during the ZTE and SPI part of the PETRA (*Acc* = 1) measurements were 60 ± 1 dBA, and 61 ± 1 dBA, respectively (Supporting Information Fig. S3).

Figure 6 presents central transverse slices of the tSNR maps at the same location of a 1000 ml cylindrical bottle phantom, comparing ZTE results from varying parameters with one reference GRE-EPI parameter set. These parameters were chosen given their relevance to the in vivo setting. The tSNR comparison indicates that a readout with Gmax=2.6 mT/m and TR/td = 2 ms / 2 µs achieves the highest tSNR. All ZTE protocols outperformed EPI, yielding up to 3-fold higher tSNR.

$O_2$- and $N_2$-saturated water had a mean image intensity of 2120±35 and 1800±58, respectively, compared to the control sample with 1980±22 (Fig. 7). $T_1$ values confirmed the difference in signal intensity with $T_1(O_{2(aq)}/N_{2(aq)}/\text{Control})$ = 883±36/1041±47/993±15 ms. T2-SPACE images exhibit a signal intensity difference of less than 2.4%, indicating a stronger T1 dependence, which the ZTE fMRI protocol could detect.

The in vivo tSNR comparison between ZTE and EPI is shown in Figure 8. For the occipital lobe, tSNR(ZTE/EPI) = 521±142/55±25 for Subject 1 , and 448±118/73±20 for Subject 2. The resulting tSNR advantage of 6-9 fold is consistent with homogeneous phantom measurements. For Subject 2, the brain extraction was imperfect. For the frontal lobe, ZTE provided a tSNR of 135±47 and 118±43 for Subjects 1 and 2, respectively.

# Discussion

In this paper, we presented the first ZTE and PETRA fMRI results in humans. Our task-fMRI experiments demonstrated robust responses in the occipital lobe, appearing more localized and less diffuse than the activation observed with EPI. Conventional GRE-EPI revealed bilateral activation patterns, whereas ZTE-based fMRI demonstrated more focal/unilateral task-related signals. While premature to draw definitive conclusions regarding their relative specificity and sensitivity, ZTE's reduced susceptibility artifacts and capacity to image regions challenging for GRE-EPI (e.g., near air-tissue interfaces) position it as a valuable complementary tool for mapping neural activity with improved spatial fidelity. The resting-state fMRI analyses, although preliminary, revealed functional networks similar to those detected with GRE-EPI. We acknowledge that the limited sample size in this initial demonstration of the ZTE fMRI prevents a comprehensive mapping of network architecture. Nevertheless, these findings provide valuable insights for future developments. Specifically, while tSNR was consistently higher in ZTE across both phantom and human brains compared to GRE-EPI, Z-scores from in vivo fMRI experiments were generally lower than those observed with EPI. This suggests that the magnitude of T1-related functional activity changes may be smaller than BOLD at 3T, highlighting the need for further optimization such as enhancing data sampling efficiency, refining hardware/sequence/reconstruction implementations, reducing temporal noise, or employing a local Tx strategy (vide infra), may be necessary before expanding to larger cohort studies to better characterize the additional networks identified with ZTE-fMRI.

3 mm nominal resolution has been used in numerous fMRI studies. Both GRE-EPI and ZTE can be implemented at higher spatial and temporal resolutions using techniques such as multi-band imaging[80,81], FOV-reduction techniques[82], or at higher magnetic fields, or by utilizing special hardware[83–85]. However, in this proof-of-principle study, a 3-mm- resolution was chosen to ensure sufficient tSNR is available. Investigations of higher spatial and temporal resolutions should be explored in future studies. The minimum achievable TE depends on the quality factor, Q, of the transmit coil, the Larmor frequency, the switching time of the electronics between transmit and receive events, and the filters in the analog-to-digital converter. In this study, a conservative TE of 60 μs was selected to ensure different RF coil setups can be run at the same TE.

In this study, the omission of SPI sampling in PETRA did not generate visible artifacts due to both the moderate resolution (3 mm³ voxels) and the short dead time ($t_d$=60 μs, $\ll T_{enc}$). However, as the dead time approaches the encoding window (i.e., $t_d \approx T_{enc}$) or as resolution increases (smaller voxel size), the k-space gap grows larger, rendering SPI sampling indispensable for artifact-free reconstructions. The adequacy of omitting SPI is thus highly specific to the protocol's resolution and timing parameters, and we caution that this approach may not be transferrable to higher-resolution or high $t_d/T_{enc}$ ratio protocols.

We also noted the limitation of the ZTE sequence in temporal resolution (3 seconds per volume). Reducing the number of spokes can improve temporal resolution but comes at the cost of signal integrity. In Supporting Information Figure S4, activation maps reconstructed from ZTE time series with 1500 and 750 spokes are compared. The distribution of activated voxels remains consistent, suggesting the potential for acceleration; however, lower T-scores for $N_{spokes}$ = 750 indicate that the k-space trajectory and reconstruction must be optimized for a reduced number of spokes. Alternative k-space sampling strategies, such as rosettes[86,87] and helical cones[88,89], have been explored in other contexts but have yet to be adapted for fMRI applications, where further improvements in spatiotemporal resolution are necessary. In this study, we intentionally used gridding reconstruction alone to better understand the properties of the ZTE sequence and provide a proper comparison against EPI, even though iterative reconstruction techniques have shown advantages in high-resolution anatomical imaging[74]. Future work should investigate the effects of regularization on the signal time course and optimize iterative reconstruction methods to achieve higher temporal resolution in ZTE-fMRI. Although for radial ZTE, trajectories including view sharing or golden-angle contradict the feature of smooth gradient transitions, the SORDINO approach can be implemented to minimize the sharp transition of gradient levels in subsequent TRs[49]. More advanced trajectories, such as rosettes[86,87,90] and helical cones,[88,89] are also directly applicable, albeit at the expense of increased acoustic noise. Helical cones can also be implemented at ultrasound frequencies to eliminate acoustic noise, similar to the high-frequency gradients used in EPI readout[91].

The activation maps derived from GRE-EPI are included primarily to illustrate expected spatial activation patterns and to demonstrate that the task paradigm was performed successfully, serving as a qualitative reference for comparison with ZTE. GRE-EPI is the gold standard in the fMRI neuroimaging community to benchmark the novel method proposed in this manuscript.

tSNR is affected not just by the instantaneous image SNR but by fluctuations over time, which depend on both thermal and physiological noise. While a reduced bandwidth decreases thermal noise, physiological or system instabilities may become more prominent at longer dwell times, limiting tSNR gains. This is observed also in phantom experiments, which might be affected by the system hardware instabilities, gradient and ADC performance, and the specifics of the reconstruction algorithm (such as gridding) which can introduce additional noise or artifacts across the time series as dwell time changes. For ZTE/PETRA sequences, the impact of dead time, gradient switching, and sampling can all modulate the noise properties in non-intuitive ways.

Although ZTE acquisitions yielded higher tSNR values compared to GRE-EPI, this increase did not translate into larger or more significant task-related activation maps, suggesting a lower temporal contrast-to-noise ratio (tCNR) in the ZTE dataset. This is likely because the inflow contribution is eliminated by the whole-body RF excitation, thus reflecting only the tissue oxygen contrast source[49]. The superior tSNR, however, suggests the potential to further improve spatiotemporal resolution, especially when tCNR is enhanced by a head-only RF excitation or a higher field magnet[49].

Variations in head size, shape, and tissue composition affect coil loading and $B_1$ uniformity, which influence both SNR and tSNR. In vivo measurements introduce physiological noise (e.g., cardiac and respiratory artifacts) that may affect EPI and PETRA differently, altering their comparative advantages. PETRA, being a 3D acquisition, and EPI, a 2D method, are differentially sensitive to these fluctuations. Even subtle differences in subject motion (including small shifts or drifts) can disproportionately degrade tSNR in EPI compared to PETRA, as EPI is more sensitive to motion-induced phase errors.

Another limitation is the acoustic noise of the proposed sequence. MB-SWIFT[62,92] and UTE-fMRI[93–95] studies have addressed susceptibility artifacts and partially mitigated the acoustic noise problems associated with EPI. Although ZTE is known to be a “silent” sequence, acquisitions with fewer spokes in dynamic imaging still result in rapid gradient change, triggering a high-pitched sound at a frequency of 1/TR. The effects of acoustic noise on resting-state networks have been reported[96,97]. ZTE-based techniques can potentially eliminate acoustic noise by minimizing changes in gradient field strength between consecutive TRs. However, considering the across-species variability in sensory sensitivity, the proposed silent encoding might not be suitable for species other than humans[98–100].

The linear relationship between T1 relaxation time and partial pressure of oxygen (pO2) has been demonstrated[101] and compared to BOLD fMRI for various applications, including tumor oxygenation assessment and prediction of radiation therapy response[102]. To the authors' knowledge, this is the first study investigating the T1-based functional changes using ZTE sequences in the human brain, which demonstrates the feasibility and functional sensitivity of ZTE-based fMRI methods at a clinical 3T system, and lays groundwork for future studies where benchmarking against broader T1-based approaches, such as PETALUTE[64]. We also acknowledge that this study alone does not directly validate sensitivity to tissue oxygenation but to changes in T1. Although preclinical studies using ZTE-based sequences[49,50] provide insight, future experiments with oxygen challenge are aimed at demonstrating the physiological basis of this contrast in humans.

To demonstrate the sensitivity of the proposed ZTE protocol to $T_1$ difference, phantom measurements using $O_2$ and $N_2$ gas solutions were performed. This simple experiment alone does not allow us to evaluate the ZTE signal intensity percent change per change in $T_1$ value: first, the $T_1$ of 993±15 ms for the reference water tube is not comparable to human brain tissue or cerebrovascular blood; second, the level of oxygen saturation might not be comparable to that expected in oxygenated blood. A more comprehensive oxygenated phantom measurement will be investigated in future studies. Here, we anticipate performing dynamic ZTE imaging on phantoms with realistic relaxation properties, constructed in high-pressure containers and infused with $O_2$ and $N_2$ gases at varying rates.

Although similar T1-weighting as ZTE can also be provided by UTE or a standard low-resolution 3D cartesian gradient echo sequence with acceleration, ZTE was preferred for its lower acoustic noise due to the continuous gradient activity with minimal ramps. Silent fMRI acquisitions may better reflect natural brain function by minimizing confounding acoustic stimulus-induced modulation. A recent study reported differences in resting-state network characteristics between continuous and periodic data acquisition approaches, with periodic scanning potentially having more disruptive effects on functional connectivity. This is particularly relevant in certain EPI studies, where long TRs are used to acquire clean electrophysiological signals between fMRI images. Supporting this perspective, animal studies using ZTE-based methods such as SORDINO[76] and MB-SWIFT[62,103,104] , offer a significant advantage, as they enable continuous data acquisition with relatively stable sound pressure levels and no audible pauses. However, auditory sensitivity varies across species, and the proposed silent encoding may need to be investigated for behavioral relevance across species.

While the use of local (brain) Tx settings has been shown to enhance functional sensitivity in rodent studies[49,62], we have chosen to address whether functional ZTE contrast can be observed in humans using a whole-body Tx configuration, without relying on inflow-related CBV contributions. The ability to detect functional activity changes without inflow enhancement is critical, given that most existing ZTE-based fMRI studies suggest that inflow is the primary source. Specifically, whole-body treatment inherently eliminates inflow effects since the blood flowing to the brain has been previously magnetized and thus does not carry higher magnetization compared to the gray matter tissue[49]. By eliminating the intrinsic inflow signal sources, the current study establishes a conservative yet encouraging lower-bound estimate of functional contrast achievable with ZTE in humans at 3T. We believe that future implementations, with the aid of a robust local Tx/Rx platform, will benefit from additional inflow-CBV contrast and further boost sensitivity, further improving the sensitivity for general fMRI applications.

# Conclusion

Our results suggest that ZTE-based fMRI can be a promising complementary approach for mapping brain function in humans, utilizing T1-related signal changes as a surrogate for neuronal activity, and offering unique advantages in specific clinical and research settings where EPI limitations are a concern.

The proposed method holds promise for researchers studying task-based functional mapping and may prove useful for measuring resting-state functional connectivity. The empirical data presented in this work demonstrate the feasibility of ZTE sequences for fMRI. We expect future coil, sequence, and reconstruction developments to address the challenges highlighted in this manuscript. Hence, these technologies may be particularly well-suited for imaging challenging populations—such as infants, patients with implants, or individuals prone to motion—and brain regions where susceptibility artifacts compromise data quality. For instance, novel techniques such as SORDINO, PETALUTE, SWIFT, looping star, etc, which aim to reduce acoustic noise, improve specificity, and increase sensitivity, have the potential to broaden the accessibility and applicability of fMRI for both preclinical and clinical scanners. While the primary focus is on neuroimaging, these technologies are readily extendable to dynamic and functional imaging of other organ systems, including the lungs, heart, liver, kidneys, and muscles.

## Acknowledgements
The first author thanks Dr. Burak Akin for teaching basics of fMRI data analysis, and Dr. Wenchao Yang for his support with early pilot experiments.

## Data and code availability
All the measurement data and analysis code will be made available upon reasonable request.

## Author contributions
ACO, MB, UE and YIS conceived and designed the study. ACO & SL collected and analyzed the data. ACO & SI developed the original XTE pulse sequence. SI, ACO and SL developed the reconstruction code. All authors provided expertise in interpreting the results and writing the manuscript.

## Competing Interests
The authors declare that they have no competing interests.

# List of Figures

**Figure 1:** Pulse sequence diagram of csPETRA sequence, which combined radial ZTE (a) and SPI (b) acquisitions to fill in the sampling gap in ZTE (c). Compressed sensing is used to reduce number of single point acquisitions (d) for fast 3D isotropic imaging.

**Figure 2:** Individual Z-score maps and time courses for ZTE and GRE-EPI data from task-based visual stimulus experiments. Slices are selected to show the peak Z-score for each subject and pulse sequence. Time courses are also shown for the peak Z-score voxels.

**Figure 3:** Group analysis of the ZTE acquisition from five healthy subjects confirm the activation in primary visual cortex and motor cortex.

**Figure 4:** The ICA components (3D volumes) for the primary visual network (PVN) and sensory motor network (SMN) are consistent between GRE-EPI and ZTE resting state acquisitions. The most informative slices are shown. All components are shown in FSL red-yellow encoding using a $3 < Z < 10$ threshold. PVN obtained from a seed-based correlation analysis is consistent with ICA for the PVN, as shown in the overlaid images: ZTE (rs-seed). Group ICA analysis of the resting state ZTE data for five volunteers confirm the location of the networks, but the spatial distributions are different due to the variations in distribution of the individual networks.

**Figure 5:** Activation maps based on image reconstructions with (PETRA) and without (ZTE) using SPI samples. Signal time courses averaged for voxels with T-scores higher than 3. There is an insignificant difference between ZTE and PETRA activation maps and the signal time courses, indicating that SPI samples could be omitted for the given acquisition parameters.

**Figure 6:** tSNR maps and tSNR values for the marked (red circle) region. (a) ZTE offers highest tSNR for TR = 2 ms and $t_d$ = 2 μs. (b) 8 μs of dwell time resulted in 50% lower tSNR (c) For TR= 1 ms, tSNR was similar to TR= 2 ms, but resulted in a higher variation within the ROI. (d) When maximum gradient strength was doubled to Gmax=10.4mT/m, tSNR dropped by 60%. (e) EPI yielded the lowest tSNR with a mean value of 240. (f) tSNR values are plotted in (g).

**Figure 7:** T2-SPACE and ZTE images of O2- and N2-enriched water tubes. ZTE fMRI sequence detects oxygenation differences due to T1 change using the same protocol as in the in vivo measurements. Signal intensity difference of up to 10% is shown compared to the control group, i.e., distilled water, for N2 and O2 dissolved water tubes.

**Figure 8:** Exemplary images comparing GRE-EPI and ZTE sequences in transverse slices, and tSNR maps obtained for two subjects. ZTE yielded higher tSNR than GRE-EPI by up to 9 fold in the occipital lobe for both subjects. Signal void artefacts due to susceptibility differences in the frontal lobe are eliminated in ZTE.

# List of Supporting Information Figures

Supporting Information Figure S1: Block design for task-based fMRI measurements.

**Supporting Information Figure S2**: Individual Z-score maps of the motor task-related activations acquired with GRE-EPI and ZTE using the 20-channel head coil. ZTE produces a more focal activation pattern, comparable to the visual task findings. Although Z-scores are generally lower with ZTE, activations consistently localize to the motor cortex..

**Supporting Information Figure S3:** Acoustic noise during PETRA with different acquisition parameters. With higher Gmax of 5.2 mT/m and 10.4 mT/m, sound levels increase with increasing acceleration factor for compressed sensing. Higher temporal resolution, i.e., lower number of spokes also increase acoustic noise due to the increased difference between te gradient waveforms between the subsequent spokes.

**Supporting Information Figure S4:** Comparison of the activation maps obtained from ZTE with $N_{spokes}$ = 1500 and 750. Consistent distribution of the focal point of the activation suggests that there is potential for higher temporal resolution. Lower T-scored, however, indicates the need for more advanced k-space trajectories and reconstruction methods.

**Supporting Information Video S1:** k-space trajectory of a single 3D volume set.

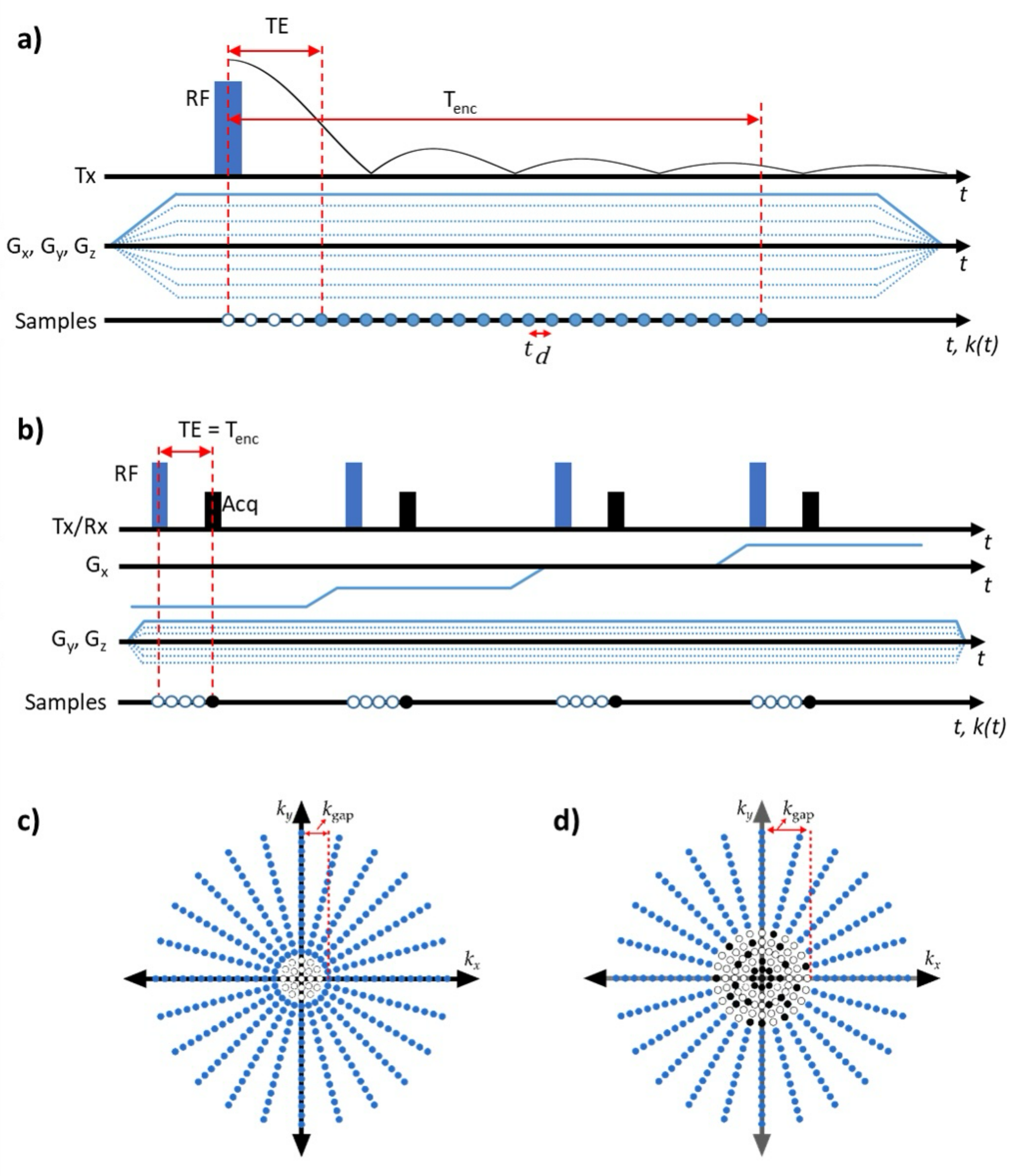
a)
TE
RF
T_enc
Tx
t
G_x, G_y, G_z
t
Samples
t_d
t, k(t)
b)
TE = T_enc
RF
Acq
Tx/Rx
t
G_x
t
t
G_y, G_z
t
Samples
t, k(t)
c)
k_y
k_gap
k_x
d)
k_y
k_gap
k_x

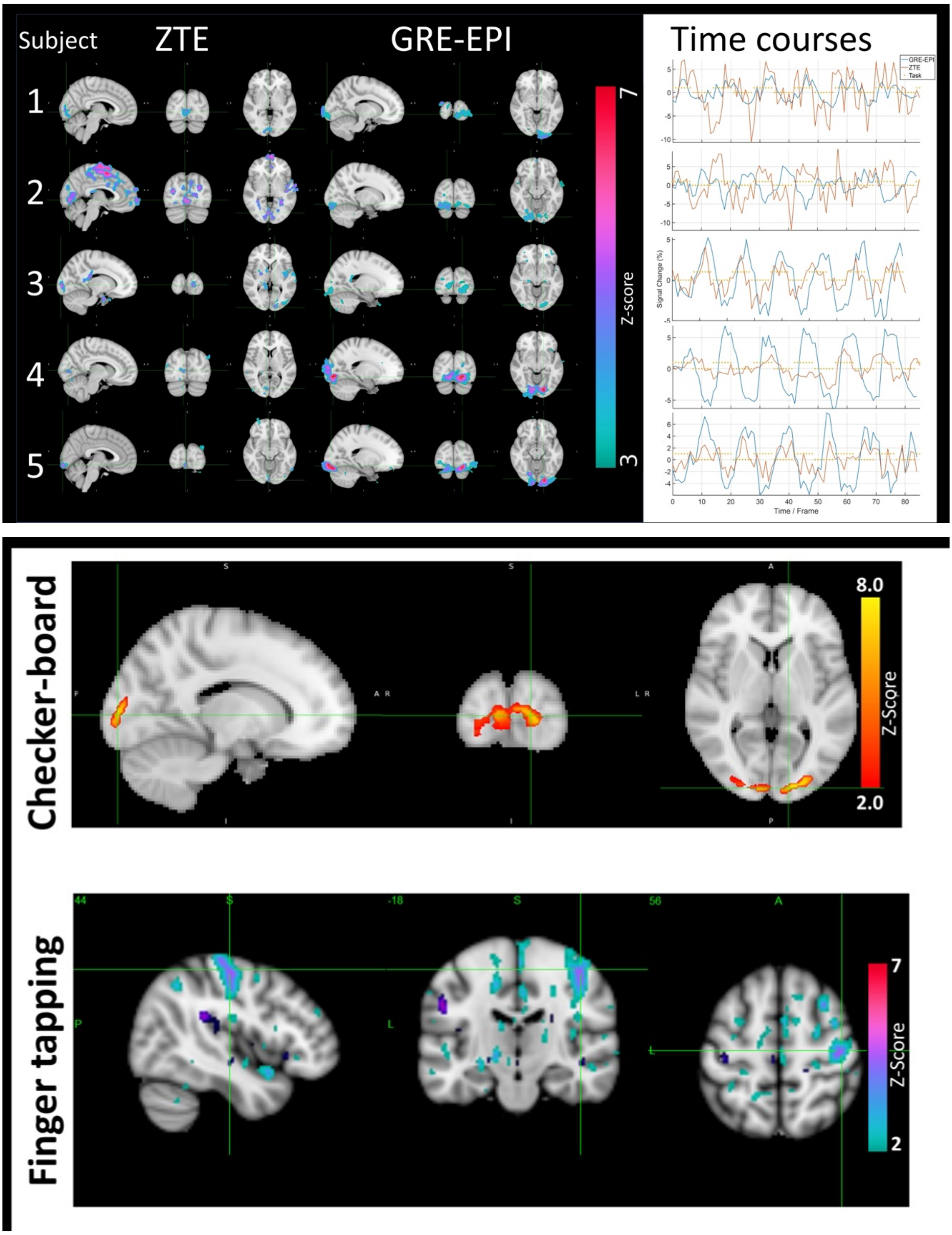
Subject
ZTE
GRE-EPI
Time courses
1
2
3
4
5
7
Z-score
3
GRE-EPI
ZTE
Task
Signal Change (%)
Time / Frame
Checker-board
8.0
Z-Score
2.0
Finger tapping
7
Z-Score
2

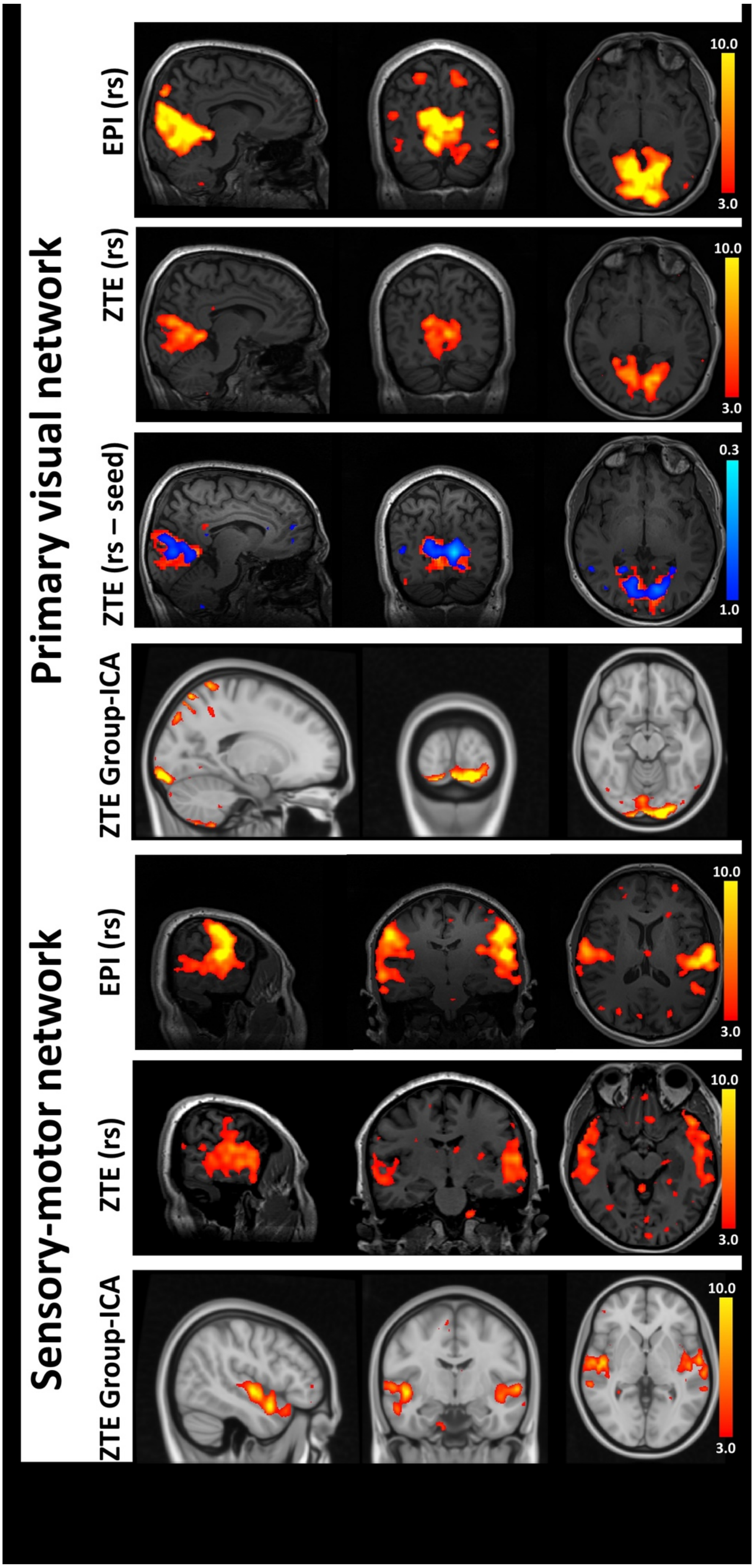

Primary visual network
EPI (rs)
ZTE (rs)
ZTE (rs – seed)
ZTE Group-ICA
Sensory-motor network
EPI (rs)
ZTE (rs)
ZTE Group-ICA
10.0
3.0
0.3
1.0

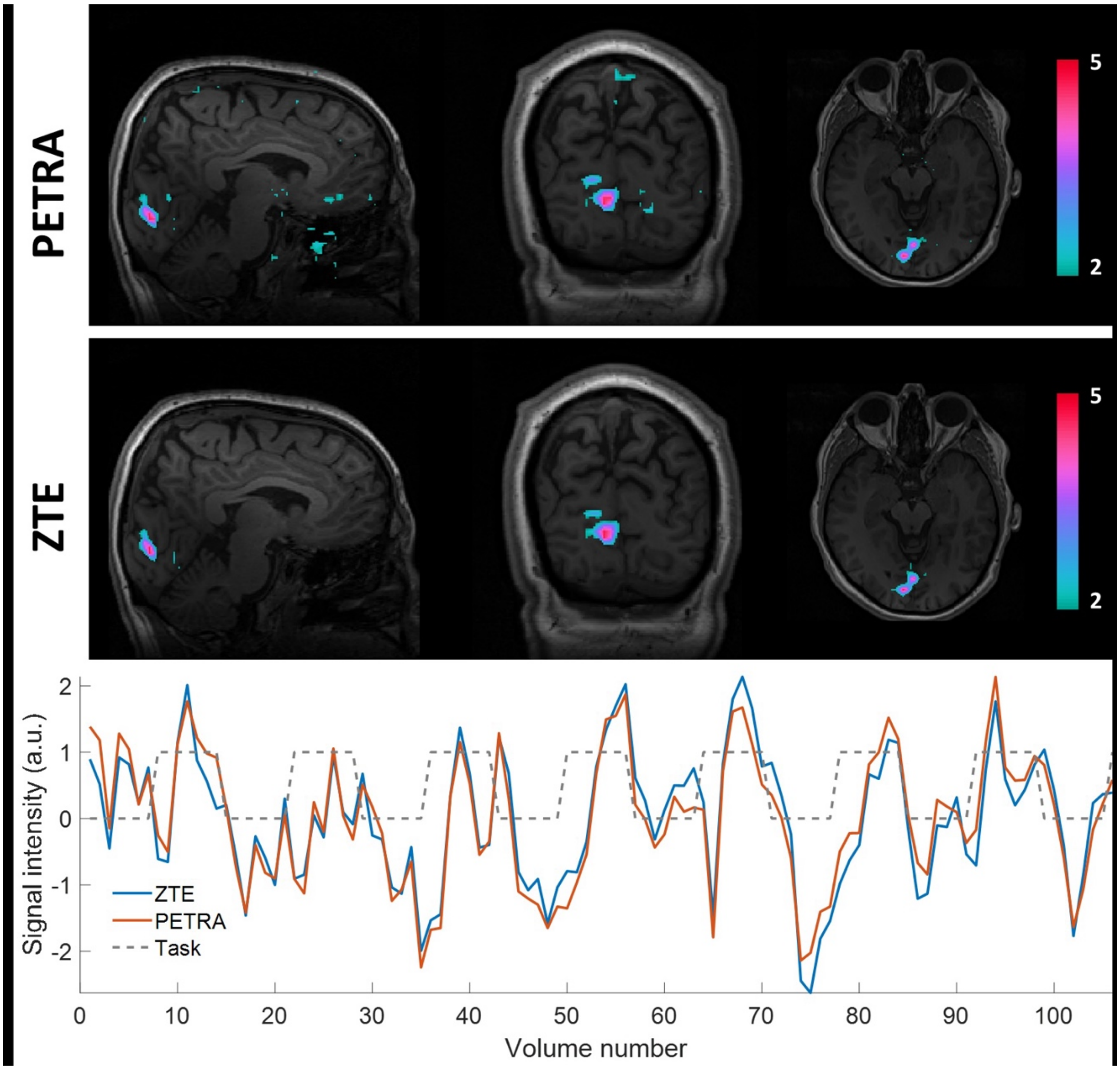
PETRA
ZTE
5
2
5
2
Signal intensity (a.u.)
Volume number
ZTE
PETRA
Task
2
1
0
-1
-2
0
10
20
30
40
50
60
70
80
90
100

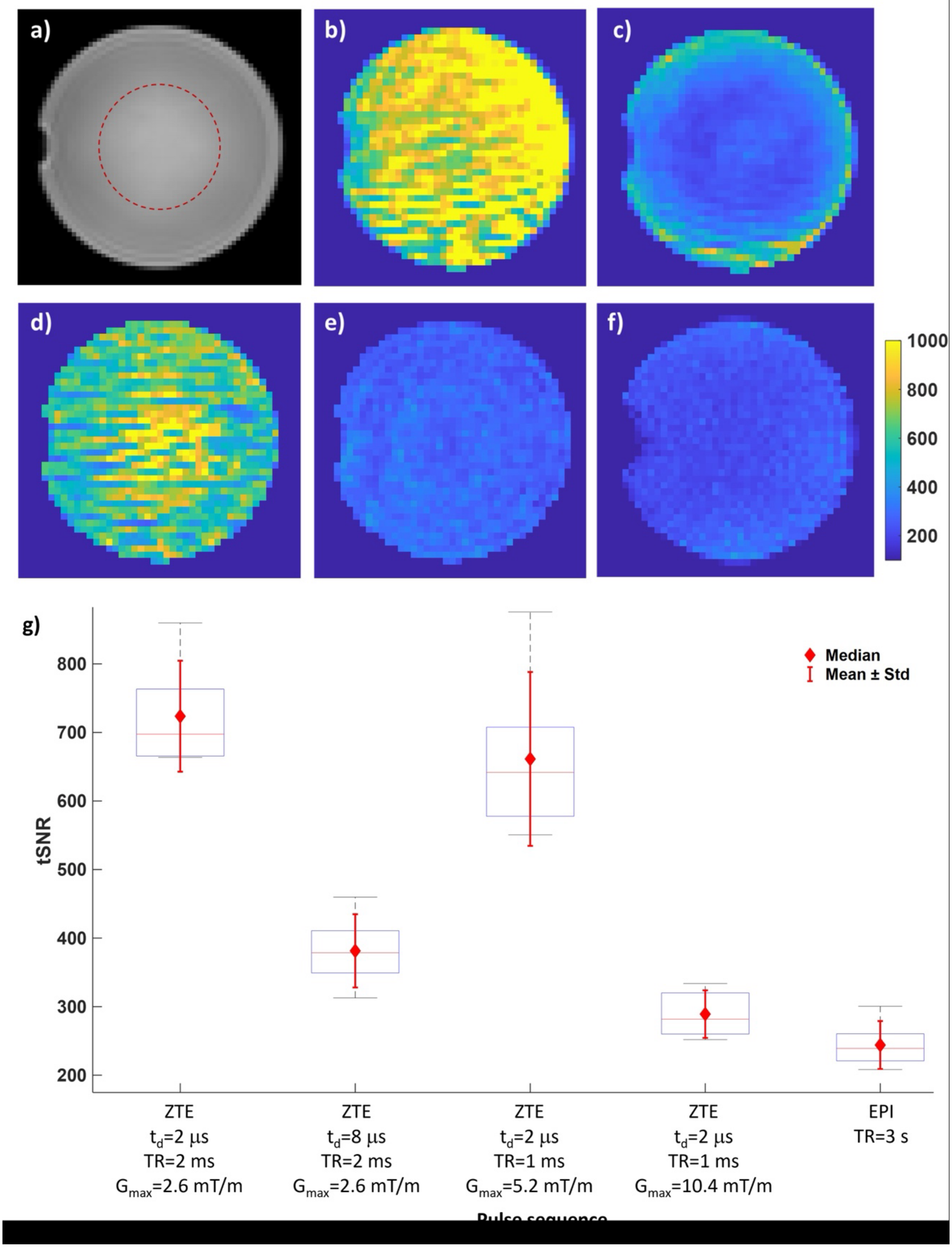
a)
b)
c)
d)
e)
f)
1000
800
600
400
200
g)
tSNR
800
700
600
500
400
300
200
Median
Mean ± Std
ZTE $t_d$=2 μs TR=2 ms $G_{max}$=2.6 mT/m
ZTE $t_d$=8 μs TR=2 ms $G_{max}$=2.6 mT/m
ZTE $t_d$=2 μs TR=1 ms $G_{max}$=5.2 mT/m
ZTE $t_d$=2 μs TR=1 ms $G_{max}$=10.4 mT/m
EPI TR=3 s

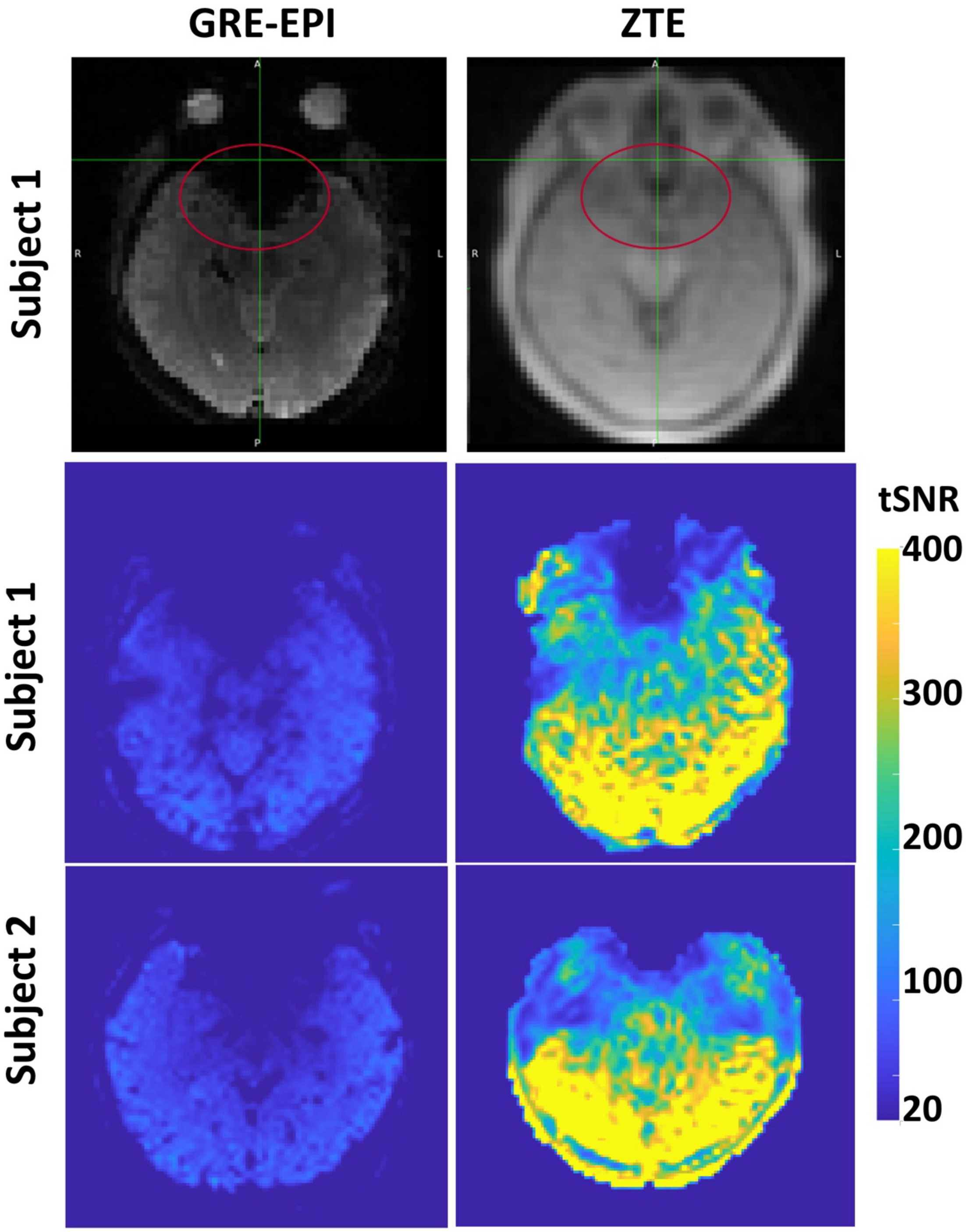
GRE-EPI
ZTE
Subject 1
Subject 1
Subject 2
tSNR
400
300
200
100
20

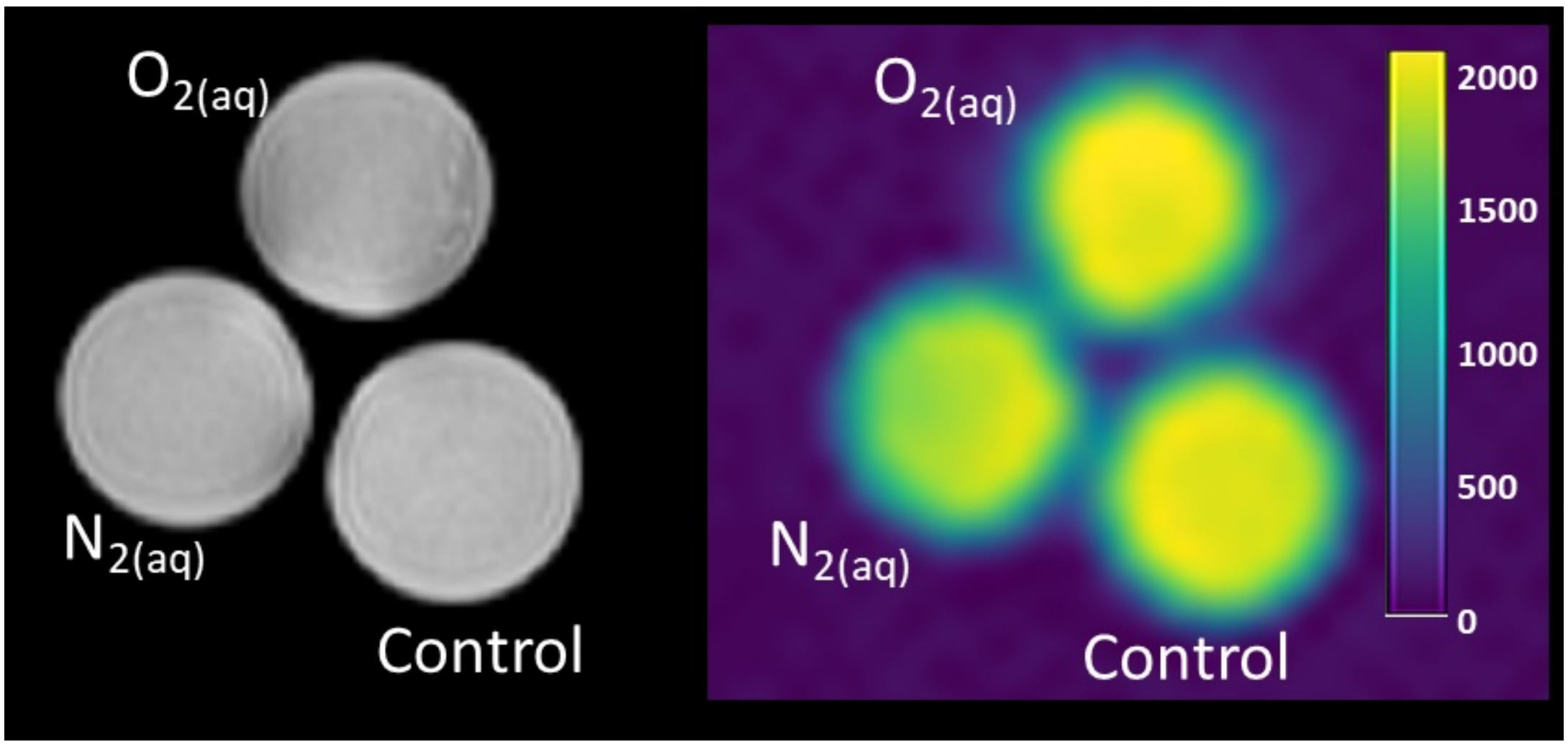

# Zero Echo Time Functional MRI in Humans

Ali Caglar Özen[1], Shuai Liu[1], Serhat Ilbey[1], Michael Bock[1], Uzay Emir[23], Yen-Yu Ian Shih[4],

[1]Division of Medical Physics, Department of Radiology University Medical Center Freiburg, University of Freiburg, Freiburg, Germany

[2] Department of Radiology, University of North Carolina, Chapel Hill, NC, USA. 3 The Lampe Joint Department of Biomedical Engineering, University of North Carolina, Chapel Hill, NC, USA.

[4] Department of Neurology, University of North Carolina, Chapel Hill, NC, USA.

Supplementary Material

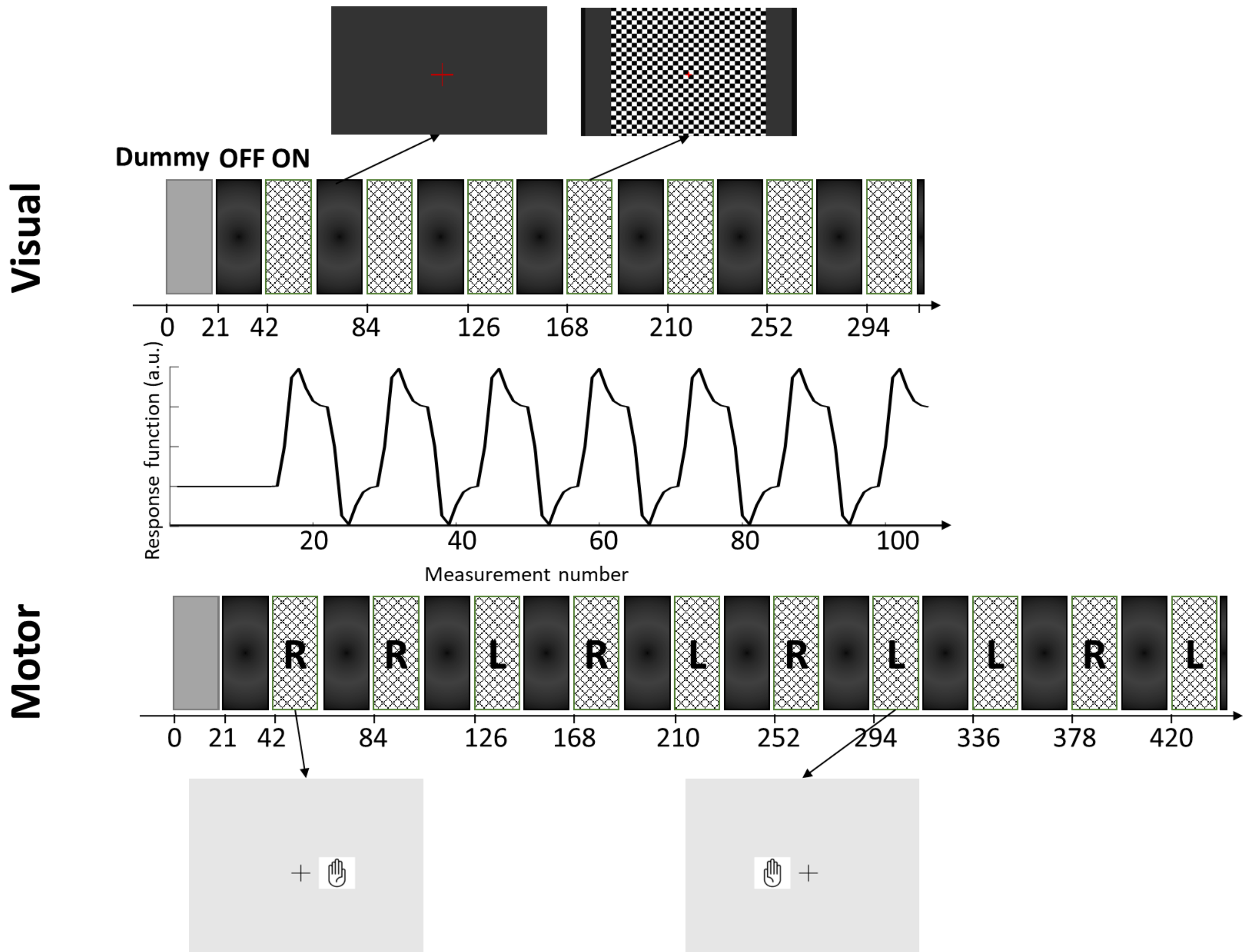

**Supporting Information Figure S1:** Block design for task-based fMRI measurements.

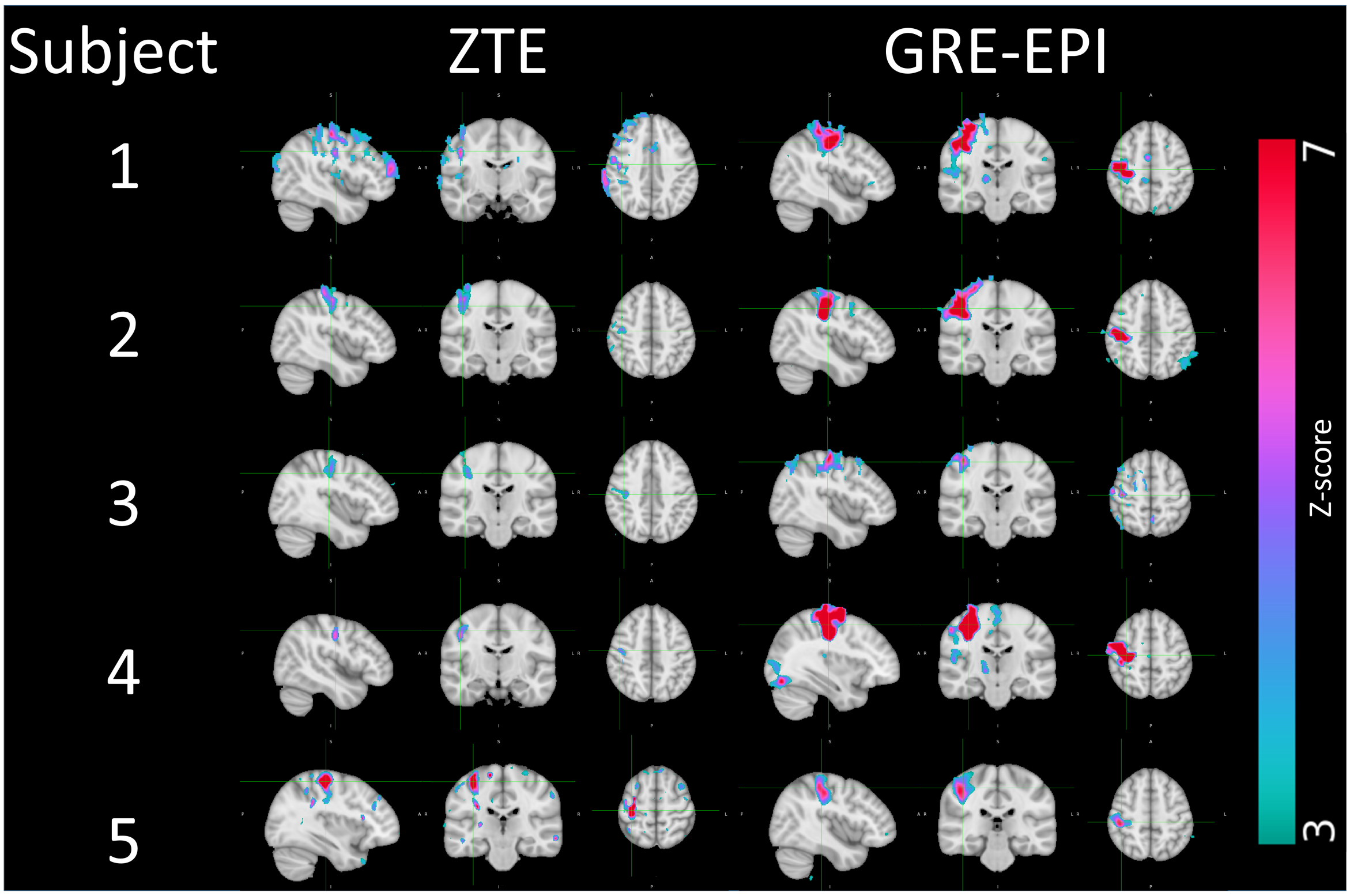

**Supporting Information Figure S2**: Individual Z-score maps of the motor task-related activations acquired with GRE-EPI and ZTE using the 20-channel head coil. ZTE produces a more focal activation pattern, comparable to the visual task findings. Although Z-scores are generally lower with ZTE, activations consistently localize to the motor cortex.

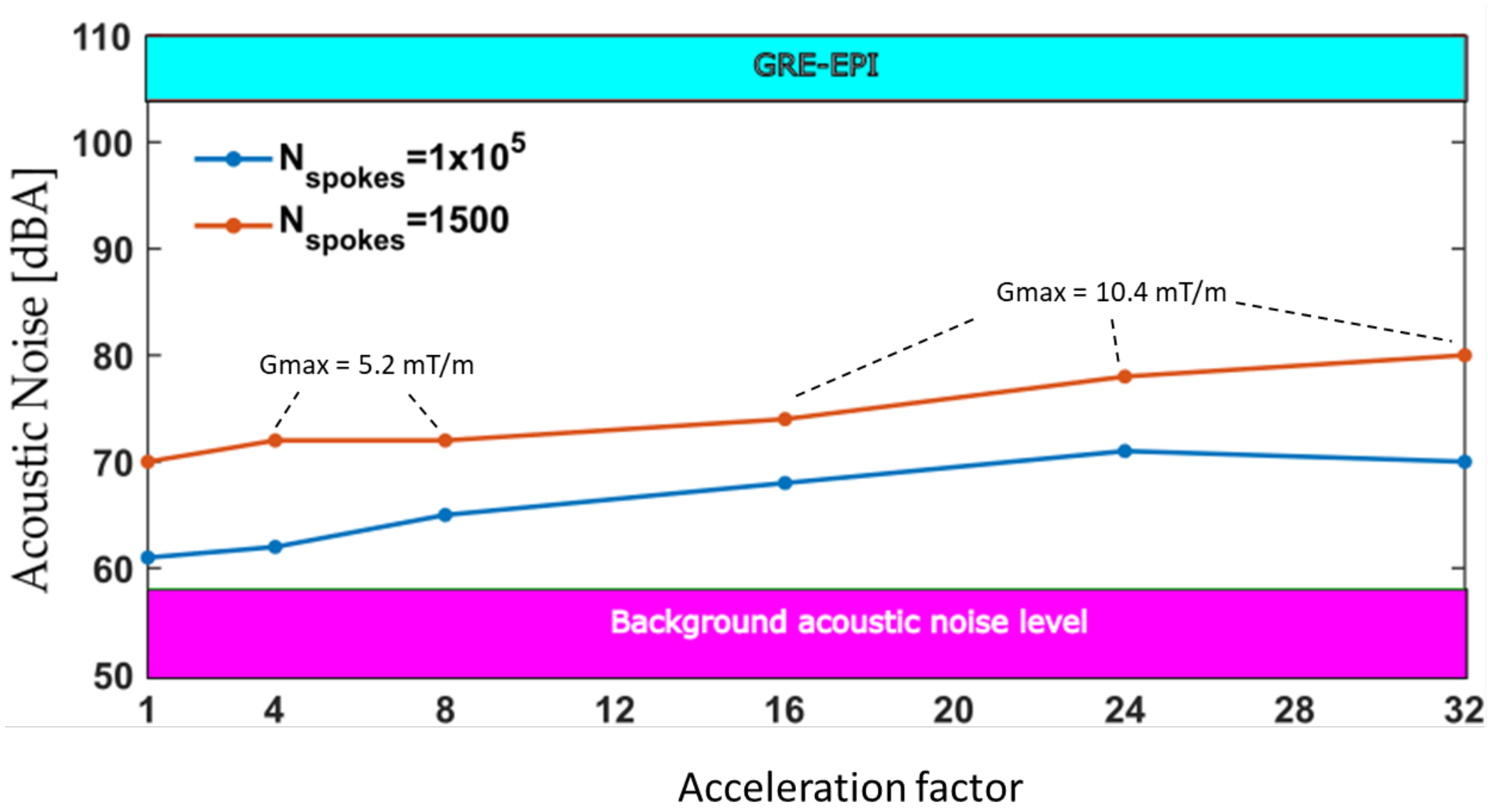

**Supporting Information Figure S3:** Acoustic noise during PETRA with different acquisition parameters. With higher Gmax of 5.2 mT/m and 10.4 mT/m, sound levels increase with increasing acceleration factor for compressed sensing. Higher temporal resolution, i.e., lower number of spokes also increase acoustic noise due to the increased difference between the gradient waveforms between the subsequent spokes.

**Supporting Information Figure S4:** Comparison of the activation maps obtained from ZTE with $N_{spokes}$ = 1500 and 750. Consistent distribution of the focal point of the activation suggests that there is potential for higher temporal resolution. Lower T-scored, however, indicates the need for more advanced

Postprocessing steps

Statistical Evaluation and Group-Level Analysis

Fsl 6.0.6 was installed in windows virtual machine Ubuntu 22.04.

A. Data Preparation

Normalization to Standard Space:
Each subject's functional and structural data were registered to the MNI152 template using nonlinear registration (FNIRT):

# Example for one subject (s1):

flirt -in s1_t1mprg_brain.nii.gz -ref MNI152_T1_2mm -omat s1_t1_to_MNI.mat

fnirt --in=s1_t1mprg_brain.nii.gz --aff=s1_t1_to_MNI.mat --config=T1_2_MNI152_2mm --iout=s1_t1_MNI.nii.gz

applywarp -i s1_vis_epi_filtered.nii.gz -r MNI152_T1_2mm -w s1_t1_to_MNI_warp -o s1_vis_epi_MNI.nii.gz

*Repeated for all subjects and functional datasets (EPI/ZTE, visual task/rest).*

B. Task-Based Activation (GLM)

1. **Merge Individual Contrasts**:

Combine all subjects' GLM output (e.g., stats_s1_zte_glm.nii.gz) into a 4D file:

fslmerge -t group_zte_glm_all.nii.gz stats_s1_zte_glm.nii.gz stats_s2_zte_glm.nii.gz ...

2. **Mixed-Effects Modeling**:

Run FLAME1 (FMRIB's Local Analysis of Mixed Effects) to assess group activation:

flameo --cope=group_zte_glm_all.nii.gz --mask=MNI152_T1_2mm_mask.nii.gz --ld=group_zte_flame --dm=design.mat --cs=group_grp.grp --runmode=flame1

design.mat: Group design matrix (e.g., one-sample t-test).

group_grp.grp: Covariate file.

3. **Multiple Comparison Correction**:

Apply cluster-based thresholding (Z > 2.3, p < 0.05):

cluster -i group_zte_flame/zstat1.nii.gz -t 2.3 –min ersize 100 -c group_zte_flame/thresh_zstat1.nii.gz -o cluster_zte.nii.gz

cluster -i groupICA_vis_output/melodic_IC.nii.gz -t 2.3 --minclustersize=100 -c ~/fsl/data/standard/MNI152_T1_2mm_gm_mask.nii.gz -o cluster_melodic_IC.nii.gz

```
n_components=$(fslval melodic_IC.nii.gz dim4)


for i in $(seq 0 $((n_components - 1))); do
  fslroi melodic_IC.nii.gz tmp_ic $i 1
  cluster -i tmp_ic.nii.gz -t 2.0 --mm --minextent=100 \
    -c ~/fsl/data/standard/MNI152_T1_2mm_gm_mask.nii.gz \
    -o cluster_melodic_IC_${i}.nii.gz
done
```

mask the original Z-map with the cluster map

```
fslmaths melodic_IC.nii.gz -mas cluster_melodic_IC.nii.gz zstat_in_clusters.nii.gz


3dClusterize -inset zstat_in_clusterICA7_vis.nii.gz -idat 0 -ithr 0 \
  -NN 2 -clust_nvox 80 -bisided -1.64 1.64 \
  -pref_map clusters.nii.gz -pref_dat cluster_zstat.nii.gz


fslroi zstat_in_clusterICA7_vis.nii.gz zstat_IC7.nii.gz 6 1
fslmaths zstat_IC7.nii.gz -mas ~/fsl/data/standard/MNI152_T1_2mm_gm_mask.nii.gz zstat_masked.nii.gz
smoothest -z zstat_masked.nii.gz -m ~/fsl/data/standard/MNI152_T1_2mm_gm_mask.nii.gz
cluster -i zstat_masked.nii.gz -t 2.3 -d 0.881491 --volume=214906 > cluster_report_with_p.txt
```

Resting-State Functional Connectivity (ICA)

Group Analysis Workflow for 5 Subjects

1. Prepare First-Level Outputs

Ensure all subjects have completed first-level GLM analysis (as previously described for individual subjects). **Required files**:

stats_s[1-5]_zte_glm.nii.gz (contrast estimates for ZTE data)

stats_s[1-5]_epi_glm.nii.gz (contrast estimates for EPI data)

2. Convert to Standard Space
Register first-level results to MNI152 template using nonlinear registration (FNIRT):

bash

# Example for s1_zte (repeat for all subjects/modalities):

applywarp -i stats_s1_zte_glm.nii.gz -r MNI152_T1_2mm -w s1_t1_to_MNI_warp.nii.gz -o stats_s1_zte_MNI.nii.gz

3. Generate Group Design Matrix

Create a design matrix (group_design.mat) for a one-sample t-test (5 subjects):

text

/NumWaves 1

/NumPoints 5

/Matrix

1

1

1

1

1

Convert to FSL format:

```
Text2Vest group_design.txt group_design.mat
```

4. Run Group-Level Analysis

```
flameo --cope=merged_contrasts.nii.gz --mask=MNI152_T1_2mm_mask.nii.gz \
    --dm=group_design.mat --tc=contrasts.con --cs=group_grp.grp \
    --runmode=flame1 --ld=group_results
```

merged_contrasts.nii.gz: 4D file combining all subjects' contrast maps (use fslmerge).

contrasts.con: Contrast file (e.g., 1 for one-sample t-test).